\documentclass{osa-article}

\journal{osac}

\articletype{Research Article}

\begin{document}

\title{Impact of cladding elements on the loss performance of hollow-core anti-resonant fibers}

\author{Md. Selim Habib,\authormark{1,*} Christos Markos,\authormark{2} and Rodrigo Amezcua-Correa\authormark{3}}

\address{\authormark{1}Department of Electrical and Computer Engineering, Florida Polytechnic University, FL-33805, USA\\
\authormark{2}DTU Fotonik, Department of Photonics Engineering, Technical Unviersity of Denmark, DK-2800, DK\\
\authormark{3}CREOL, The College of Optics 
 and Photonics, University of Central Florida, FL-33816, USA}

\email{\authormark{*}mhabib@floridapoly.edu} 



\begin{abstract}
Understanding the impact of the cladding tube structure on the overall guiding performance is crucial for designing single-mode, wide-band, and ultra low-loss nested hollow-core anti-resonant fiber (HC-ARF). Here we thoroughly investigate on how the propagation loss is affected by the nested elements when their geometry is realistic (\emph{i.e.,} non-ideal). Interestingly, it was found that the size rather than the shape of the nested elements, have a dominant role in the final loss performance of the regular nested HC-ARFs. We identify a unique `V-shape' pattern for suppression of higher-order modes loss by optimizing free design parameters of HC-ARF. 
We find that a 5-tube nested HC-ARF has wider transmission window and better single-mode operation than 6-tube HC-ARF. We show that the propagation loss can be significantly improved by using \emph{anisotropic} nested anti-resonant tubes elongated in the radial direction. Our simulations indicate that with this novel fiber design, a propagation loss as low as 0.11 dB/km at 1.55 $\mu$m can be achieved. Our results provide design insight toward fully exploiting single-mode, wide-band, and ultra low-loss HC-ARF. In addition, the extraordinary optical properties of the proposed fiber can be beneficial for several applications such as future optical communication system, high energy light transport, extreme non-nonlinear optics and beyond. 
\end{abstract}

\section{Introduction}

Hollow-core anti-resonant fibers (HC-ARFs) (also called negative curvature HC-ARFs or tubular HC-ARFs) have attracted massive interest in the fiber optics community due to their extraordinary optical properties including low latency, non-linearity, power overlap with glass, and anomalous dispersion over a wide transmission range~\cite{pryamikov2011demonstration,belardi2014hollow_nested,poletti2014nested,habib2019single,sakr2019ultrawide,sakr2020interband,bradley2018record,bradley2019antiresonant,debord2013hypocycloid,debord2017ultralow,yu2016negative,yu2012low,van2016modal,gao2018hollow,jasion2020hollow,habib2016low}. One of the unique features of HC-ARFs is that in the anti-resonant condition,  $>$99.99$\%$ of light can be guided in the air-core and a tiny fraction of light leaks towards cladding which offers very high damage threshold  and reduced material attenuation~\cite{poletti2014nested,habib2019single}. The remarkable optical properties of HC-ARF find various applications  such as short reach data transmission~\cite{sakr2020interband,sakr2019ultrawide}, long distance data transmission \cite{nespola2020transmission}, next generation scientific instruments and polarization purity~\cite{Taranta2020Exceptional}, high power delivery~\cite{michieletto2016hollow,gebhardt2017nonlinear}, gas-based nonlinear optics~\cite{travers2011ultrafast,russell2014hollow,adamu2019deep,adamu2020noise,habib2017soliton,wang2020high,markos2017hybrid}, extreme UV light generation~\cite{habib2019extreme,habib2018multi}, non linear microendoscopy~\cite{kudlinski2020double}, mid-IR transmission~\cite{kolyadin2013light,urich2013flexible}, optofluidic \cite{hao2018optimized}, and terahertz applications~\cite{anthony2011thz,hasanuzzaman2015low,sultanan2020exploring}. 

The light guiding mechanism of HC-ARF is based on inhibited coupling (IC) between the core-guided modes and the continuum of modes of the cladding \cite{pearce2007models}, including also anti-resonant effect~\cite{poletti2014nested}. The idea of IC guiding mechanism was first introduced by Couny \emph{et. al.,}~\cite{couny2007generation} which can be explained by the high degree of the transverse-field mismatch between the core and cladding modes (CMs). Moreover, the coupling between CMs and core mode can be strongly inhibited by using a negative curvature (or hypocycloid) core contour~\cite{debord2013hypocycloid} and carefully playing with the cladding tubes~\cite{habib2019single}. This remarkable guiding mechanism of light in HC-ARF offers much wider transmission window and low dispersion compared to the first generation of hollow-core fiber so called hollow-core photonic bandgap fibers (HC-PBGFs) in which light guides based on photonic bandgap effect~\cite{cregan1999single,amezcua2008control}. 

In recent years, intense efforts have been dedicated towards designing low-loss HC-ARFs as an alternative optical transmission medium by several research groups ~\cite{pryamikov2011demonstration,belardi2014hollow_nested,poletti2014nested,habib2019single,debord2013hypocycloid,debord2017ultralow, bradley2018record,sakr2019ultrawide,sakr2020interband,yu2016negative,bradley2019antiresonant}. All of the fiber designs proposed so far are based on `negative curvature' anti-resonant tubes. Most of the fiber designs are relying upon tubes with circular \cite{debord2017ultralow}, ``ice--cream cone'' shape \cite{yu2012low,yu2016negative,van2016modal}, conjoined  \cite{gao2018hollow},  elliptical \cite{habib2016low}, semi--circular \cite{habib2015low}, nested  \cite{belardi2014hollow_nested,poletti2014nested,habib2019single}, split cladding \cite{huang2016hollow}, and so on. Early design of HC-ARFs comprise of single ring anti-resonant tube consisting of either 6, 7, 8, or more tubes used in the cladding which defines the negative curvature \cite{pryamikov2011demonstration,michieletto2016hollow,debord2017ultralow,belardi2014hollow}. The lowest loss of 7.7 dB/km at 750 nm was experimentally reported in a single ring structure HC-ARF \cite{debord2017ultralow}. Despite of promising loss values found in this simple geometry, it is still far from the optical losses that the conventional telecommunication optical fibers offer today \cite{gao2018hollow}. In addition, single ring HC-AR fiber typically has high bend loss because CMs and core mode can be coupled under small bend radius \cite{habib2015low}. 

One of the most efficient ways to reduce propagation loss and bend loss of HC-ARF is to add small tubes (also called `nested tubes') inside outer tubes which was first proposed in 2014 \cite{belardi2014hollow_nested}. Addition of the so-called nested tube offers a significant enhancement in the IC between CMs and core mode \cite{poletti2014nested,habib2019single}, and thus reduces propagation loss and bend loss. Recently, considerable improvements in nested HC-ARFs have been achieved and some outstanding results were reported in telecommunication window. For example, a propagation loss of 1.3 dB/km at 1450 nm with 65 nm bandwidth below 1.5 dB/km was achieved with a 6-tube nested HC-ARF \cite{bradley2018record}. Most recently, a record loss of 0.28 dB/km from 1510 nm -- 1600 nm has been experimentally demonstrated \cite{jasion2020hollow}. This incredible nested HC-ARF development opens a new paradigm for investigating short reach optical communication. However, the impact of nested tubes on propagation loss and higher-order modes (HOMs) loss are not fully exploited and few issues still remain challenging such as: (i) what will be the impact on overall loss performance when nested tubes of HC-ARF become non-ideal rather than a perfect circle, and (ii) how to reduce the propagation loss close to standard single-mode fiber (SSMF) with wide-band transmission window, and truly single-mode behavior that have all favourable properties for future optical communication system.

The aim of this work is to investigate the influence of anti-resonant tubes of HC-ARF on the overall loss performance, in particular when the nested tubes \emph{swerve} from its ideal structure. Controlling the shape of the nested tubes to be circular (ideal nested tubes) \cite{sakr2020interband}, and maintaining them in a particular position \cite{bradley2018record} are typically crucial during the fabrication process. The nested tubes become non-ideal (we call them `realistic' in this work) and how much deviations will be allowed to maintain low-loss. We find that when the nested tube approaches to more realistic scenario, almost similar propagation loss can be obtained by making slightly larger nested tubes than the ideal case. In addition, we propose a 5-tube nested HC-ARF fiber design which exhibits propagation loss $<$0.25 dB/km, wide transmission window, and truly effectively single-mode operation. Finally, we investigate the effect of anisotropic nested tubes, elongated in the radial direction on propagation loss. The propagation loss significantly improved by applying anisotropic nested tubes in which leakage loss has negligible contribution. A propagation loss as low as 0.11 dB/km is achieved at 1.55 $\mu$m. Our results presented in this work will provide better understanding in designing ultra low-loss, wide-band, and single-mode HC-ARFs. 

\begin{figure}
  \begin{center}
  \includegraphics[width=4.5in]{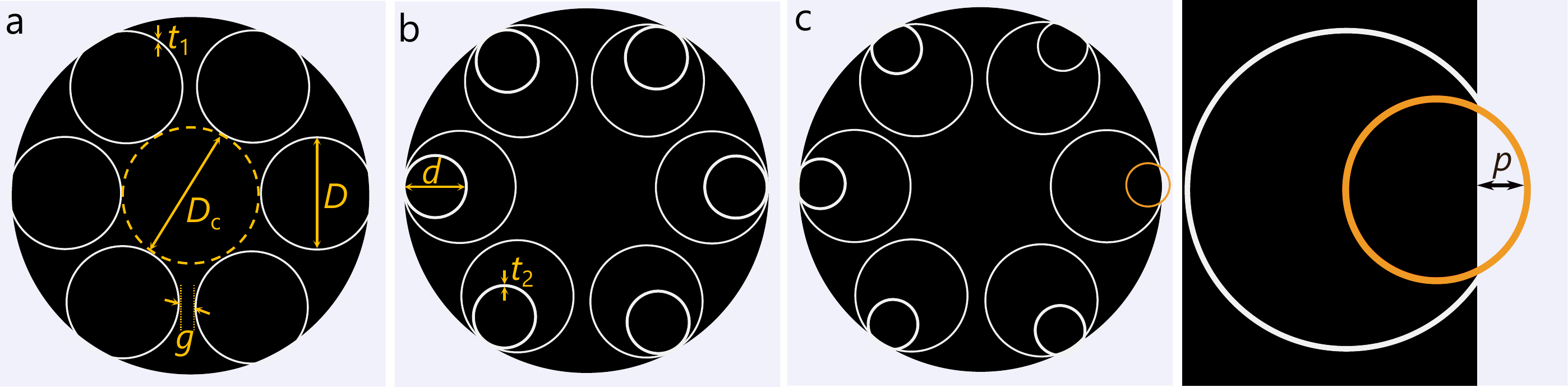}\\
  \caption{HC-ARF geometries considered in the simulations. (a) Typical HC-ARF design with six circular tubes; (b) nested HC-ARF; (c) realistic nested HC-ARF in which nested tubes are penetrated toward the outer jacket forming non-ideal tubes. The magnification of one of the nested tubes is shown next to (c). The penetration of nested tube is indicated by $p$. All fibers have the same core diameter $D_\text{c}$ = 35 $\mu$m, wall thickness of outer and inner tubes $t_1$/$t_2$ = 415/460 nm, and $g$ = separation between the outer tubes. The design parameters are similar to \cite{sakr2019ultrawide,sakr2020interband}.}\label{fig:fig_1}
  \end{center}
\end{figure}

\section{Fiber geometry}
Figure~\ref{fig:fig_1} shows the different HC-ARF designs considered in our investigations. The fiber design parameters are similar to recently reported fiber designs \cite{sakr2019ultrawide,sakr2020interband} which aim to provide a low-loss transmission window in the telecommunication band. The fiber parameters are chosen to operate in the first anti-resonant pass band because it offers larger bandwidth than higher anti-resonant pass band \cite{sakr2020interband}. Figure~\ref{fig:fig_1}(a) shows a typical 6-tube HC-ARF design of core diameter $D_\text{c}$, tube diameter $D$, wall thickness $t_{1}$, and a gap separation between the outer tubes, $g$. The gap separation provides a node-free core-boundary which offers better loss properties and flatter transmission window \cite{poletti2014nested,habib2019single}. An \emph{ideal} nested HC-ARF of outer diameter $D$, inner tube diameter $d$, and wall thickness of outer and inner tubes $t_1/t_2$ is presented in Fig.~\ref{fig:fig_1}(b). We choose different wall thickness of outer and inner tubes to reflect a more realistic case than an ideal one \cite{sakr2020interband}. Finally, a realistic nested HC-ARF is shown in Fig.~\ref{fig:fig_1}(c). In this fiber design, the nested tubes penetrate towards the outer jacket tube forming non-circular shape tubes. This type of fiber structures are typically common during the fabrication process \cite{sakr2019ultrawide,sakr2020interband,bradley2018record,bradley2019antiresonant}. In our numerical calculations, we choose a core diameter of $D_\text{c}$ = 35 $\mu$m, and silica wall thickness of outer and inner tubes of $t_1/t_2$ = 415/460 nm unless otherwise stated. The outer diameter $D$ is related to the core diameter $D_\text{c}$, wall thickness $t_1$, and number to tubes $N$, which can be written as \cite{wei2017negative},
\begin{equation}
\begin{split}
D = \frac{\frac{D_\text{c}}{2}sin(\frac{\pi}{N})-\frac{g}{2}-t{_1}(1-sin(\frac{\pi}{N}))}{1-sin(\frac{\pi}{N})},
\end{split}
\label{eq:eq1}
\end{equation}
In our simulations, a small penetration of $t_1$ of all cladding tubes into the outer silica tube and bend diameter of 32 cm (fiber is typically spooled to a 100 cm circumference bobbin \cite{bradley2018record}) was considered in order to achieve well agreement with the practical case \cite{belardi2014hollow}. We found that the loss remains almost constant over a wide range of penetration thickness. 

\section{Numerical results and discussions}

To perform the numerical simulations, we used finite-element modeling based on COMSOL software. A perfectly-matched layer (PML) boundary was placed outside the fiber domain to accurately calculate the fiber modal properties \cite{poletti2014nested,habib2019single}. The mesh size and PML boundary conditions were optimized according to \cite{sakr2020interband,sakr2019ultrawide,poletti2014nested,habib2019single}. The calculations were performed using extremely fine mesh sizes of $\lambda/6$ and $\lambda/4$ in the silica walls and air regions respectively \cite{poletti2014nested}, which give excellent agreement with experimental results \cite{poletti2014nested,hayes2017antiresonant}. In our propagation loss calculations, all possible loss contributions were considered \emph{i.e.,} leakage loss (also know as confinement loss), effective material loss (EML), surface scattering loss (SSL), and micro-bend loss unless otherwise stated. The SSL was calculated using the method reported in \cite{roberts2005ultimate,poletti2014nested}. It is to be noted that SSL is much lower in HC-ARF compared to HC-PBGF because the light interaction with glass is quite low \cite{poletti2014nested}. The micro-bend loss was calculated using the approach demonstrated in \cite{fokoua2016microbending,sakr2020interband} and the micro-bend loss can be written as: $\alpha_\text{bend} = C(\Delta\beta)\beta_{0}(|<0|x^2|0>|^2-|<0|x^2|1>|^2)$; where $C(\Delta\beta)$ is the power spectral density, $\beta_0$ = propagation constant of the fundamental mode, and $|x|$ is the spot size. In order to estimate EML, we calculate the power overlap with silica walls, and then added to other loss contributions. The material attenuation of silica was taken from \cite{humbach1996analysis}. The calculated power overlap with silica walls for all calculations were less than $10^{-4}$.

\begin{figure}
  \begin{center}
  \includegraphics[width=3.7in]{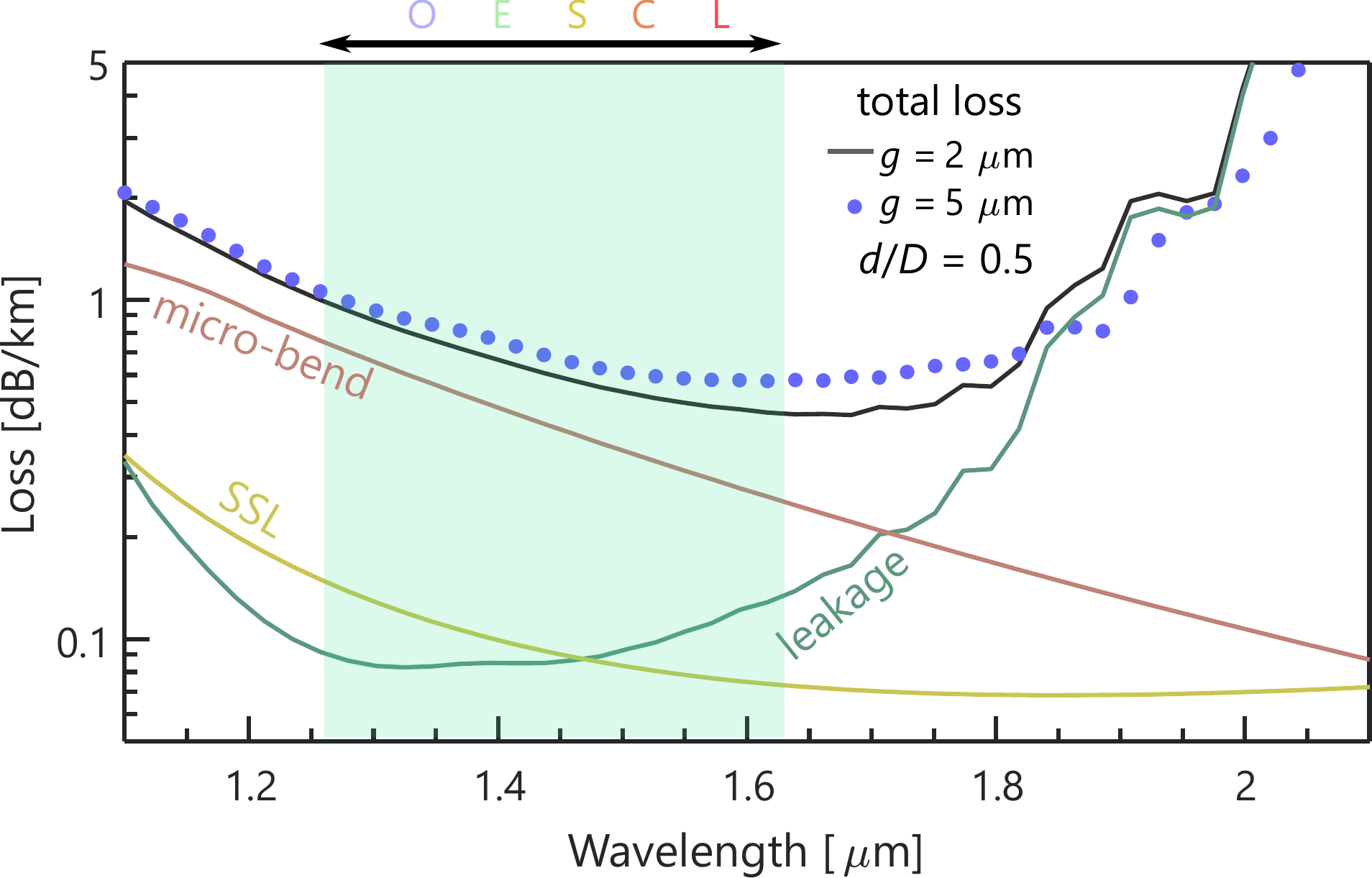}\\
  \caption{Calculated propagation loss spectra of ideal nested HC-ARFs for two  gap separations $g$ = 2 $\mu$m (black solid line) and $g$ = 5 $\mu$m (violet closed circle) with $d/D$ = 0.5. The HC-ARF has core diameter, $D_\text{c}$ = 35 $\mu$m and average silica wall thickness, $t_1/t_2$ = 415/460 nm. The $O+E+S+C+L$ band (1260 nm $-$ 1625 nm) is shown in light cyan color bar inside the figure.}\label{fig:fig_2}
  \end{center}
\end{figure}

\begin{figure}
  \begin{center}
  \includegraphics[width=3.7in]{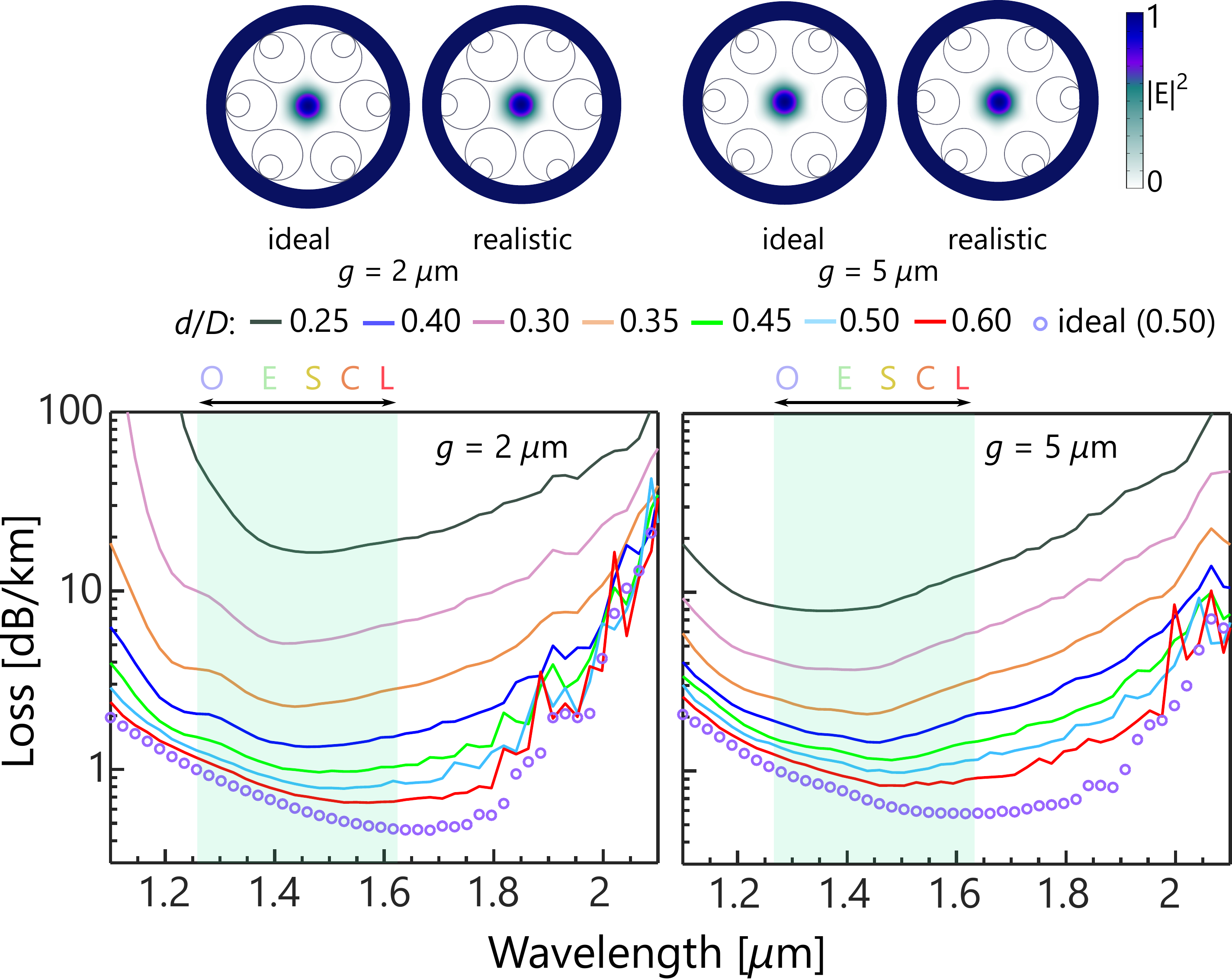}\\
  \caption{Calculated propagation loss spectra of ideal and realistic nested HC-ARFs for two  gap separations $g$ = 2 $\mu$m (left) and $g$ = 5 $\mu$m (right) and different $d/D$. For ideal HC-ARF, $d/D$ = 0.5. All fiber designs have the same core diameter, $D_\text{c}$ = 35 $\mu$m and uniform silica wall thickness, $t_1/t_2$ = 415/460 nm. The electric field intensities LP$_{01}$-like mode for different gap separation and $d/D$ at 1.55 $\mu$m are shown at the top on a linear color scale.}\label{fig:fig_3}
  \end{center}
\end{figure}

\begin{figure}[ht]
  \begin{center}
  \includegraphics[width=3.7in]{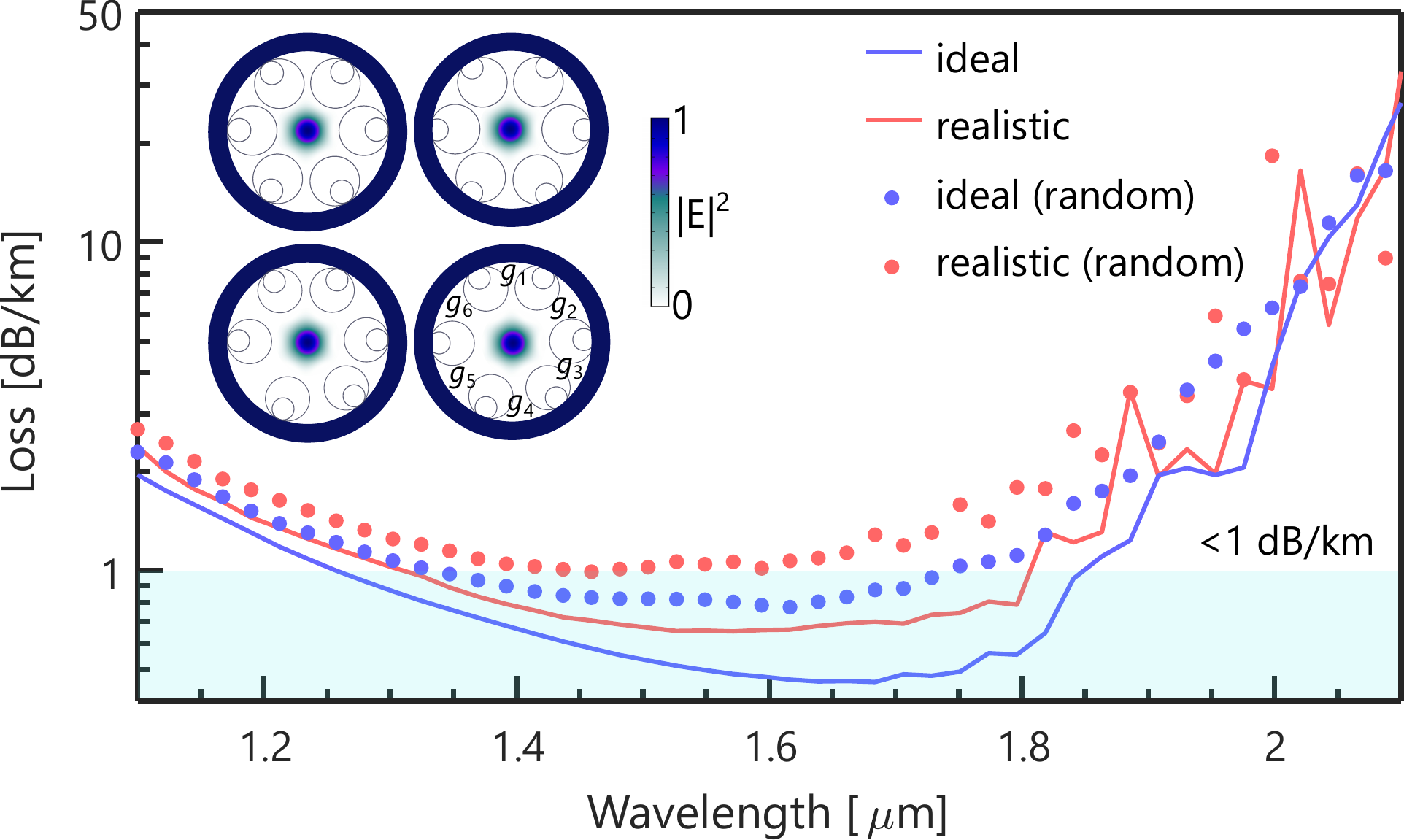}\\
  \caption{Calculated loss spectra of ideal and realistic nested HC-ARF. For ideal and realistic HC-ARF $d/D$ = 0.5 and 0.65 were chosen respectively. All fiber designs have the same core diameter, $D_\text{c}$ = 35 $\mu$m and average silica wall thickness, $t_1/t_2$ = 415/460 nm. The solid lines and closed circles present a fixed gap separation of $g$ = 2 $\mu$m, and random gap separation respectively.  The random gap separations was chosen $\{\it{g}_{1},\it{g}_2,\it{g}_3,\it{g}_4,\it{g}_5,\it{g}_6\}$ = \{3.4,7.51,2.36,6.48,7,9.04\} $\mu$m. The electric field intensities of LP$_{01}$-like mode for different designs at 1.55 $\mu$m is shown in the inset on a linear color scale.}\label{fig:fig_4}
  \end{center}
\end{figure}

\subsection{Ideal 6-tube nested HC-ARF}
The calculated propagation loss spectra of an ideal 6-tube HC-ARF for a fixed normalized nested tube ratio of $d/D$ = 0.5, and gap separations of $g$ = 2 $\mu$m and 5 $\mu$m, is illustrated in Fig.~\ref{fig:fig_2}. The micro-bend dominates over leakage loss in the short wavelength regime whereas it dominates over SSL in the entire wavelength regime. The loss profiles for both gap separations are comparable. It means micro-bend loss is crucial and can not be neglected for designing HC-ARF $<$1 dB/km for this particular fiber specifications. More details on how to reduce the micro-bend loss is explained in \emph{Sec. 3.6}. In the shorter wavelength regime, both have similar loss profile. The smaller gap separation has better loss performance in the mid-band wavelength regime. The gap separation $g$ = 5 $\mu$m has slightly wider bandwidth than $g$ = 2 $\mu$m. 

\subsection{Effect of gap separation, $g$ and normalized nested tube ratio, $d/D$ on loss}
Here, we investigate the effect of gap separation, $g$ and normalized nested tube ratio, $d/D$ on the propagation loss for both ideal and realistic HC-ARF designs. Nested tube penetration of $p$ = 3 $\mu$m was chosen for realistic HC-ARF. The offset of the nested tubes is set to be larger than the outer tubes. This is because the nested tubes are smaller and tend to penetrate more into the outer jacket under normal fabrication conditions whereas the outer tubes seem be pushed less into the cladding (more details can be found in \cite{sakr2020interband}). In our analyses, we consider small and moderately large gap separations of $g$ = 2 $\mu$m and 5 $\mu$m respectively. Different $d/D$ was chosen for realistic design whereas $d/D$ of 0.5 was considered for an ideal HC-ARF. The propagation loss and normalized electric field intensities are illustrated in Fig.~\ref{fig:fig_3}. The solid lines and open circles correspond to the loss of realistic and ideal HC-ARF respectively. The optimized HC-ARF has a loss of $<$2 dB/km from 1260 nm to 2000 nm (740 nm) with $<$0.7 dB/km loss at 1550 nm. A few loss peak oscillations were observed in the longer wavelengths for larger $d/D$. When $d/D$ increases, two glass nodes will form with the outer tubes which is close to the core boundary giving rise to additional Fano resonances at the longer wavelengths \cite{gao2018hollow}. 
These oscillations of the loss peaks can be reduced by carefully choosing $d/D$. It can be seen from Fig.~\ref{fig:fig_3} that the propagation loss strongly depends on $d/D$ and $g$. The loss increases significantly for small $d/D$, whereas loss decreases with the increase of $d/D$. The impact of $d/D$ on propagation loss is relatively low for larger $d/D$ in the range of 0.4$-$0.6.  It is worth mentioning that to obtain a comparable loss spectra, a realistic fiber design requires slightly larger $d/D$ than ideal nested HC-ARF. When designing an HC-ARF $<$1 dB/km, choosing $d/D$ is one of the key factors whereas the shape of the nested tube weakly depends on the propagation loss. 
Controlling the shape of relatively thinner nested tubes during fabrication process is crucial \cite{sakr2019ultrawide,sakr2020interband}. 
Therefore, the above finite-element modeling results provide important evidence on whether an ideal nested tube is essential for designing ultra-loss fiber. The finite-element modeling predicts that ideal nested tubes can be avoided and replaced with more realistic nested tubes (\emph{i.e.,} semi-circular tubes) with slightly higher $d/D$.
These investigations are important from a fiber fabrication point of view as it can relax some fabrication constraints for nested HC-ARFs.

\begin{figure}[t]
  \begin{center}
  \includegraphics[width=3.5in]{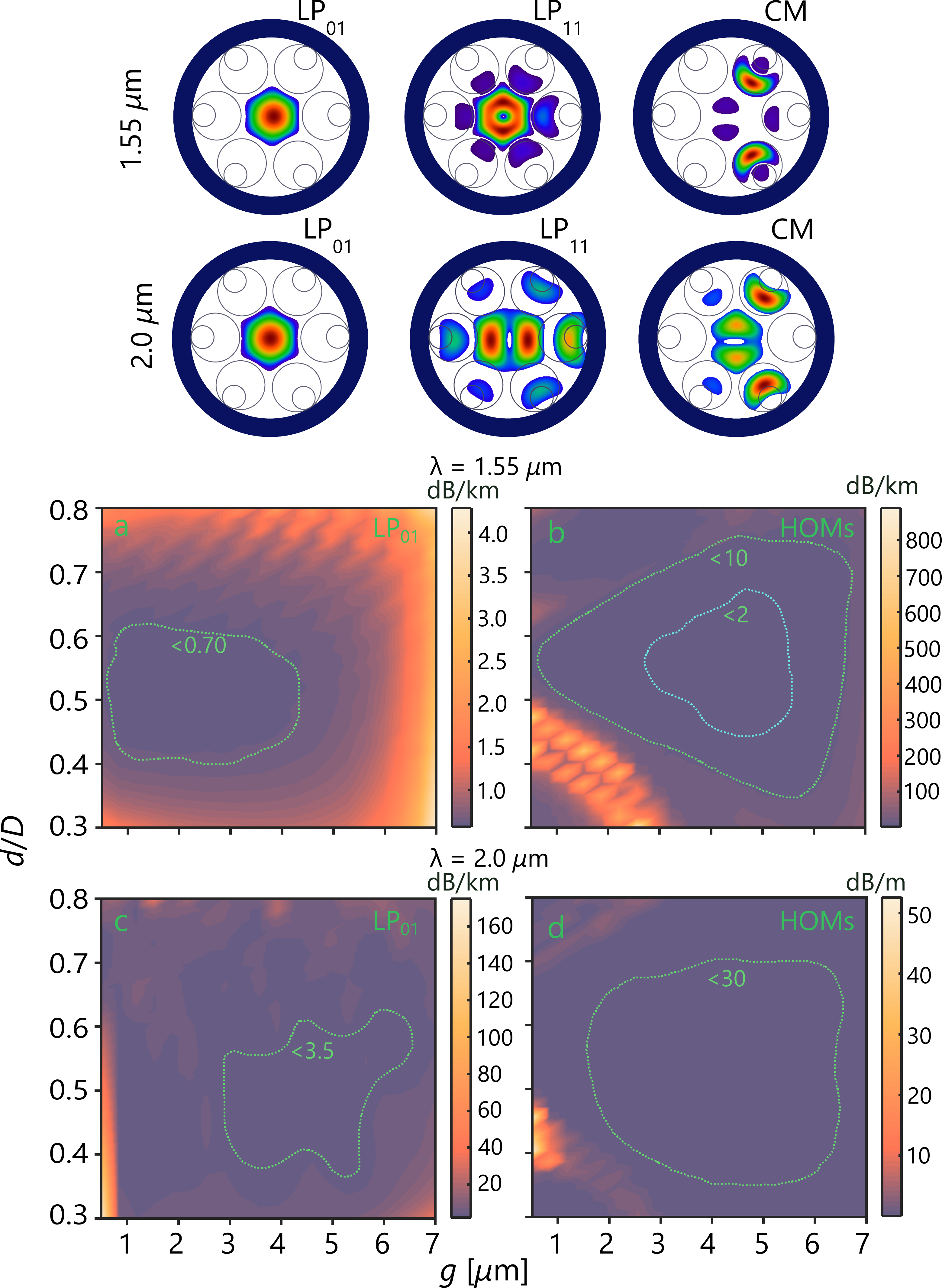}\\
  \caption{Calculated propagation loss of (a,c) LP$_{01}$-like FM, and (b,d) HOMs as a function of $d/D$ with different values of gap separation,  $g$. HOM loss in (b,d) is defined as the lowest loss among the four LP modes (LP$_{11a}$, LP$_{11b}$, LP$_{21a}$, and LP$_{21b}$). All simulations are performed at 1.55 $\mu$m and 2.0 $\mu$m. The fiber has a fixed core diameter, $D_\text{c}$ = 35 $\mu$m and average silica wall thickness, $t_1/t_2$ = 415/460 nm. To plot the 2D surface plots, gap separation, $g$, and normalized nested tube ratio, $d/D$ were scanned with 20 and 30 data points respectively and between the data points are interpolated. The mode-field profiles of LP$_{01}$, LP$_{11}$, and CM are shown for 1.55 $\mu$m and 2 $\mu$m at the top.}\label{fig:fig_5}
  \end{center}
\end{figure}
\begin{figure}
  \begin{center}
  \includegraphics[width=3.6in]{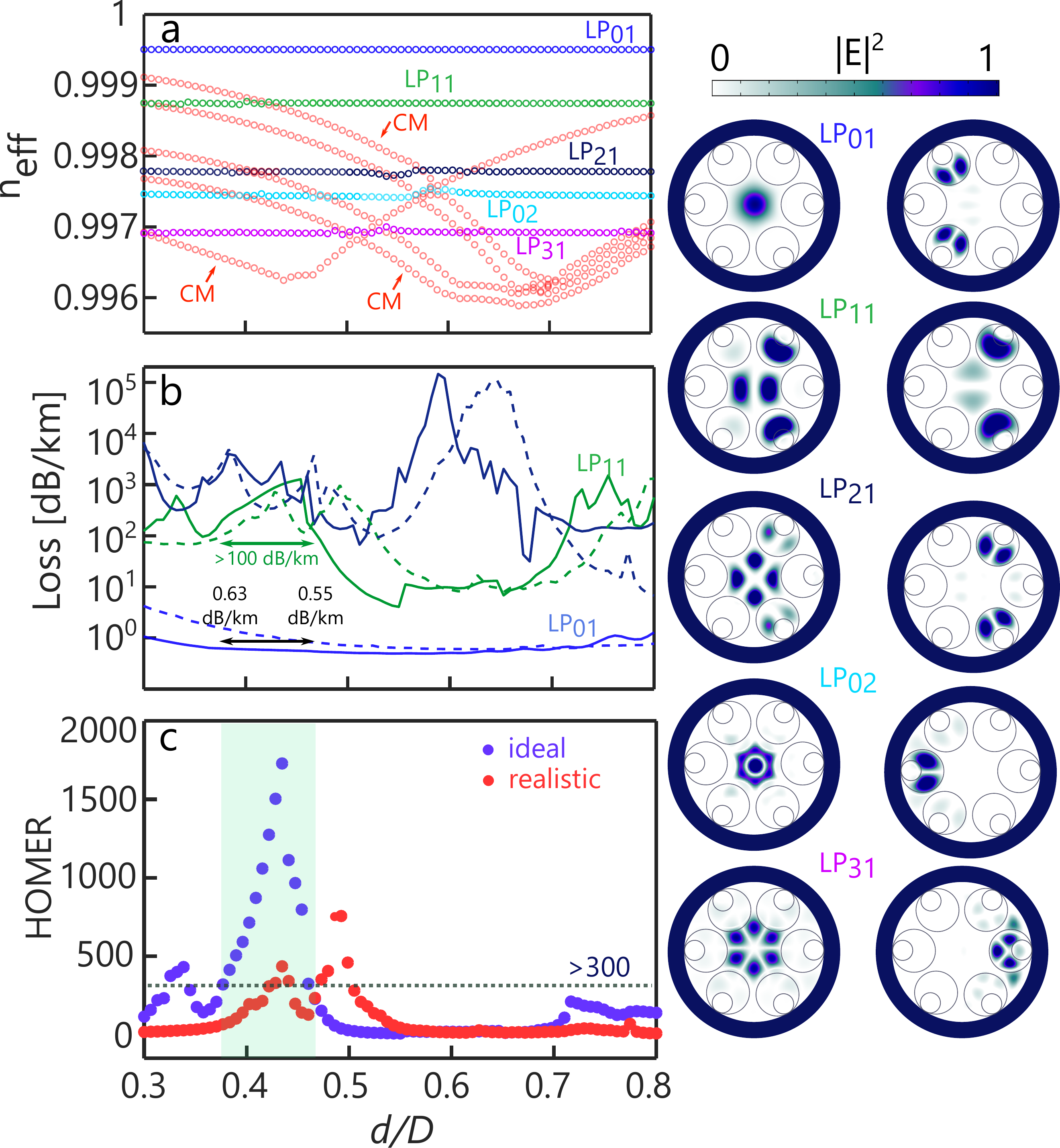}\\
  \caption{Effect of changing normalized nested tube ratio, $d/D$ on (a) effective refractive index, (b) loss, and (c) HOMER with a fixed core diameter, $D_\text{c}$ = 35 $\mu$m, average wall thickness, $t_1/t_2$ = 415/460 nm, and gap separation, $g$ = 2 $\mu$m. The normalized tube ratio, $d/D$ were scanned from 0.3 to 0.8. The simulations were performed at 1.55 $\mu$m. The electric field intensities of the first five core-guided modes (left) and CMs (right) are shown for $d/D$ $\approx$ 0.43 and $g$ = 2 $\mu$m on a linear color scale for ideal HC-ARF. The highlighted light cyan bar shows the range of $d/D$ where HOMER $>$300. The solid and broken lines present propagation loss of ideal and realistic HC-ARF respectively in (b). For realistic HC-ARF, nested tube penetration of $p$ = 3 $\mu$m was chosen.}\label{fig:fig_6}
  \end{center}
\end{figure}

\subsection{Effect of random separation of anti-resonant tubes}
Here, we discuss the effect of random separation of the anti-resonant tubes on the propagation loss. This type of random separation of tubes typically arises for thinner tubes. The results are illustrated in Fig.~\ref{fig:fig_4}. The propagation loss for random separation is higher as expected than a fixed gap separation of $g$ = 2 $\mu$m. The effect is more pronounced in the mid-band wavelength regime compared to short and long wavelength regime. However, despite of such random change of anti-resonant tubes, loss remains sub-dB/km in the wavelength range of 1400 nm$-$1600 nm (200 nm) which indicates the robustness of the fiber. A uniform shift of the position of nested tubes alone might appear during the fabrication and such a shift of nested tubes from its original position is usual \cite{bradley2018record,belardi2015design}. In reality the shift of the tubes would be random from tube to tube, but a uniform shift of the tubes can give an indication of the fabrication tolerances. The HC-ARF is typically robust with such a shift of the nested tubes for a wide range of position angle, $\theta$ according to \cite{habib2015low,belardi2015design,jasion2019novel} (see Ref. \cite{habib2015low} for the definition of the position angle, $\theta$), and we expect to get similar results.

\subsection{HOMs suppression in HC-ARF}
In this section, we discuss how the free design parameters (\emph{i.e.,} nested tube ratio, $d/D$, and gap sepation, $g$) of a 6-tube HC-ARF can be optimized to suppress the higher order modes (HOMs). Figure~\ref{fig:fig_5} depicts the 2D surface plots of fundamental mode (FM) and HOMs propagation loss as a function of $g$ and $d/D$ for $\lambda$ = 1.55 $\mu$m and 2 $\mu$m. These 2D maps are useful tool to identify the range of low-loss FM and effectively single-mode operation region.
The optimization of low-loss and effectively single-mode region were obtained by systematically changing the gap separation, $g$, and normalized tube ratio, $d/D$. It can be seen from Fig.~\ref{fig:fig_5}(a) that at $\lambda$ = 1.55 $\mu$m the FM loss remains $<$0.7 dB/km in the range of 0.5 $\mu$m $<g< $ 4.9 $\mu$m and 0.4 $<d/D< $0.55. The FM loss increases when $g>$5.5 $\mu$m irrespective of $d/D$.  The loss of HOMs (lowest loss among the LP$_{11a}$, LP$_{11b}$, LP$_{21a}$, and LP$_{21b}$ modes) is shown in Fig.~\ref{fig:fig_5}(b). By properly choosing $g$ = 2.55 $\mu$m and $d/D$ = 0.3, the HOM loss can be made $>$0.9 dB/m while maintaining the FM loss slightly higher than 1 dB/km. A comparatively similar trend was obtained at longer wavelength of 2 $\mu$m. The FM loss remains $<$4 dB/km for a wide range of $g$ and $d/D$. A HOM loss of $>$50 dB/m is found for small gap separation of $g$ = 0.5 $\mu$m and $d/D$ = 0.45. These large loss value of HOMs for both cases are due to the strong coupling between HOMs and CMs \cite{poletti2014nested,habib2016low,habib2019single}. The results suggest that by properly selecting $g$ and $d/D$, HOMs can be significantly suppressed leading to effectively single-mode fiber. The higher-order-mode extinction ratio (HOMER), which is defined as the ratio between the propagation loss of the HOM with the lowest loss and the propagation loss of FM is used to identify single-modeness of the fiber \cite{poletti2014nested}. At $\lambda$ = 1.55 $\mu$m, HOMER as high as $>$1000 is obtained whereas almost 5 times higher HOMER $>$5000 is obtained for $\lambda$ = 2 $\mu$m. 
The reason of higher HOMER at longer wavelengths is because the phase matching conditions between HOMs with CMs can be relatively easier to occur at longer wavelengths where the light is not strongly confined. In addition, the core-guided HOMs are more easier to be coupled to the cladding as the wavelength increases. This can be also seen from the mode-field profiles for the case of $\lambda$ = 1.55 $\mu$m, and $\lambda$ = 2 $\mu$m.

In order to get better understanding of how $g$ and $d/D$ affects the FM loss, HOMs loss, and HOMER, we further investigate with a fixed core diameter, $D_c$ = 35 $\mu$m, $g$ = 2 $\mu$m, $t_1/t_2$ = 415/460 nm while we change $d/D$ from 0.3 to 0.8. It also assists to identify the phase matching of HOMs with CMs. Figure~\ref{fig:fig_6} shows the effective refractive index, $n_{\rm{eff}}$  of the first five core-guided modes and CMs, propagation loss of the first three core-guided modes (LP$_{01}$, LP$_{11}$, LP$_{21}$), and HOMER as a function of $d/D$. The loss of HOMs \emph{i.e.,} LP$_{02}$ and LP$_{31}$ are not shown because these modes have higher losses compared to LP$_{11}$ and LP$_{21}$. The effective refractive index of the core-guided modes remain almost constant as a function of $d/D$ while it significantly changes for CMs. For FM (LP$_{01}$)-like core-guided mode, effective refractive index is far away from the CMs avoiding any phase matching (anti-crossing) between them which ensures low-loss for FM. However, strong phase matching between the core-guided HOMs and CMs occur for various locations of $d/D$. Due to this strong phase matching between the core-guided HOMs and CMs, HOMs typically have higher losses. The HOMER as high as $>$300 can be made in the range of $d/D$ = 0.38$-$0.47 for ideal HC-ARF. However, the peak of the HOMs and HOMER slightly shifts for realistic the realistic HC-ARF, and also has low HOMER compared to ideal HC-ARF. 

\begin{figure}
  \begin{center}
  \includegraphics[width=3.6in]{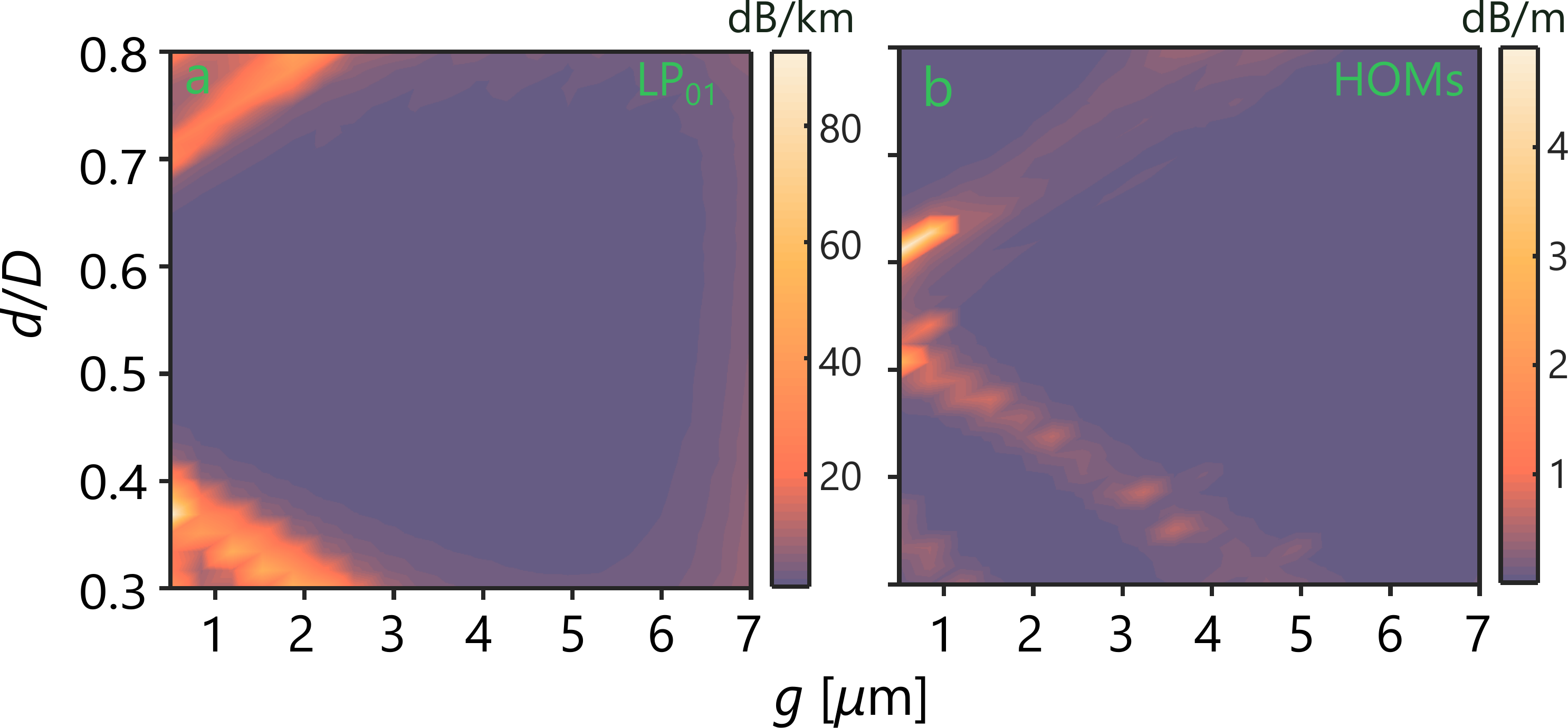}\\
  \caption{Calculated bend loss of (a) LP$_{01}$-like FM and (b) HOMs (right) as a function of $d/D$ with different values of gap separation, $g$ for 5 cm bend radius. HOMs loss is defined as the lowest loss among the four LP modes (LP$_{11a}$, LP$_{11b}$, LP$_{21a}$, and LP$_{21b}$). All simulations are performed at 1.55 $\mu$m. The fiber has a fixed core diameter, $D_\text{c}$ = 35 $\mu$m and average silica wall thickness, $t_1/t_2$ = 415/460 nm. To plot the 2D surface plots, gap separation, $g$, and normalized nested tube ratio, $d/D$ were scanned with 20 and 30 data points respectively and between the data points are interpolated. The colorbar for LP$_{01}$ and HOMs are in dB/km and dB/m respectively. }\label{fig:fig_7}
  \end{center}
\end{figure}

\begin{figure}[t]
  \begin{center}
  \includegraphics[width=3.4in]{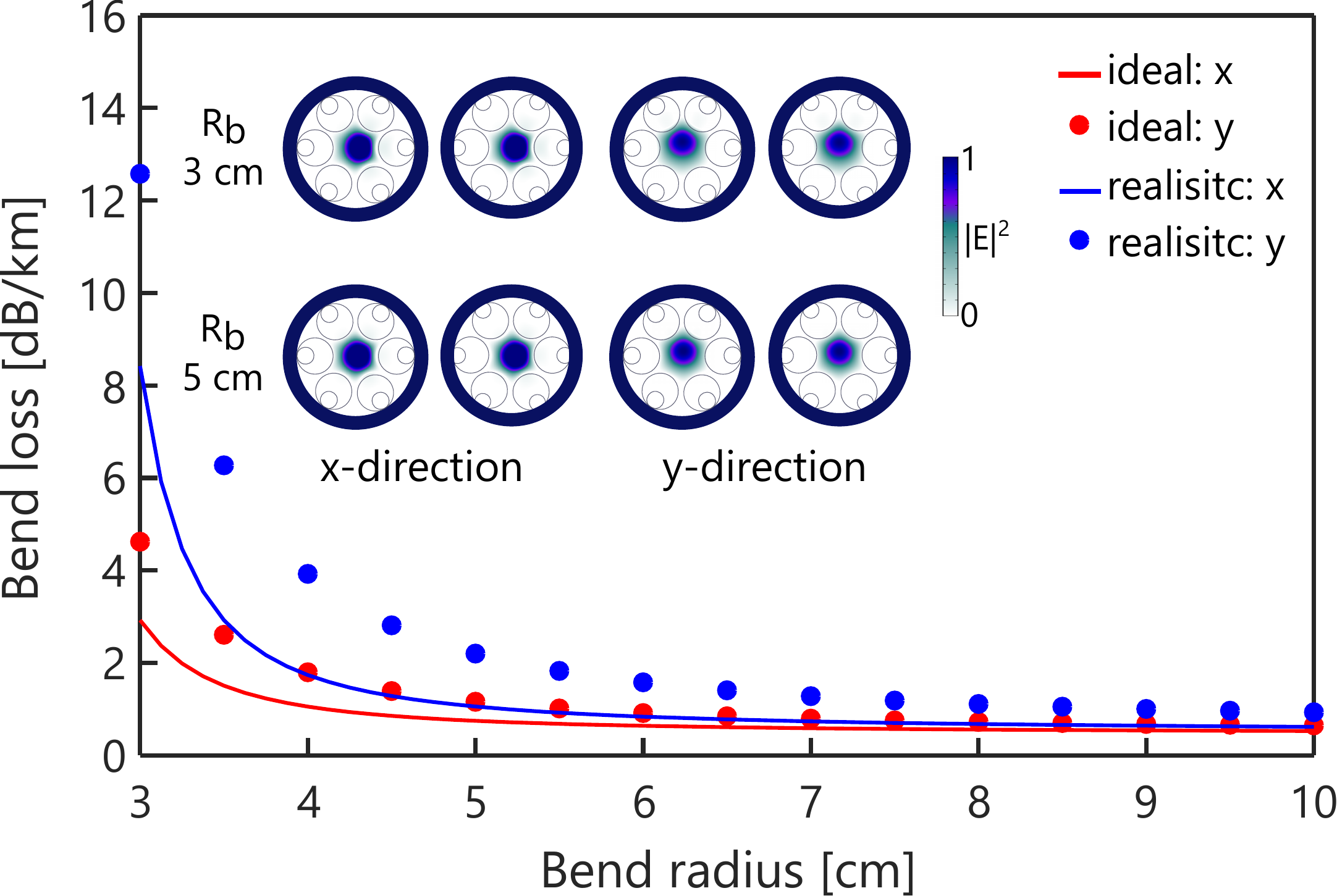}\\
  \caption{Calculated bend loss of 6-tube nested ideal and realistic HC-ARF as a function of bend radius.  The fiber has a fixed core diameter, $D_{c}$ = 35 $\mu$m and average silica wall thickness, $t_1/t_2$ = 415/460 nm, $g$ = 2 $\mu$m. Ideal and realistic fiber has $d/D$ of 0.5 and 0.6. The electric field intensities of ideal and realistic design for bend radius of 3 cm and 5 cm are shown in the inset on a linear color scale.}\label{fig:fig_8}
  \end{center}
\end{figure}

\begin{figure}[t]
  \begin{center}
    \includegraphics[width=3.8in]{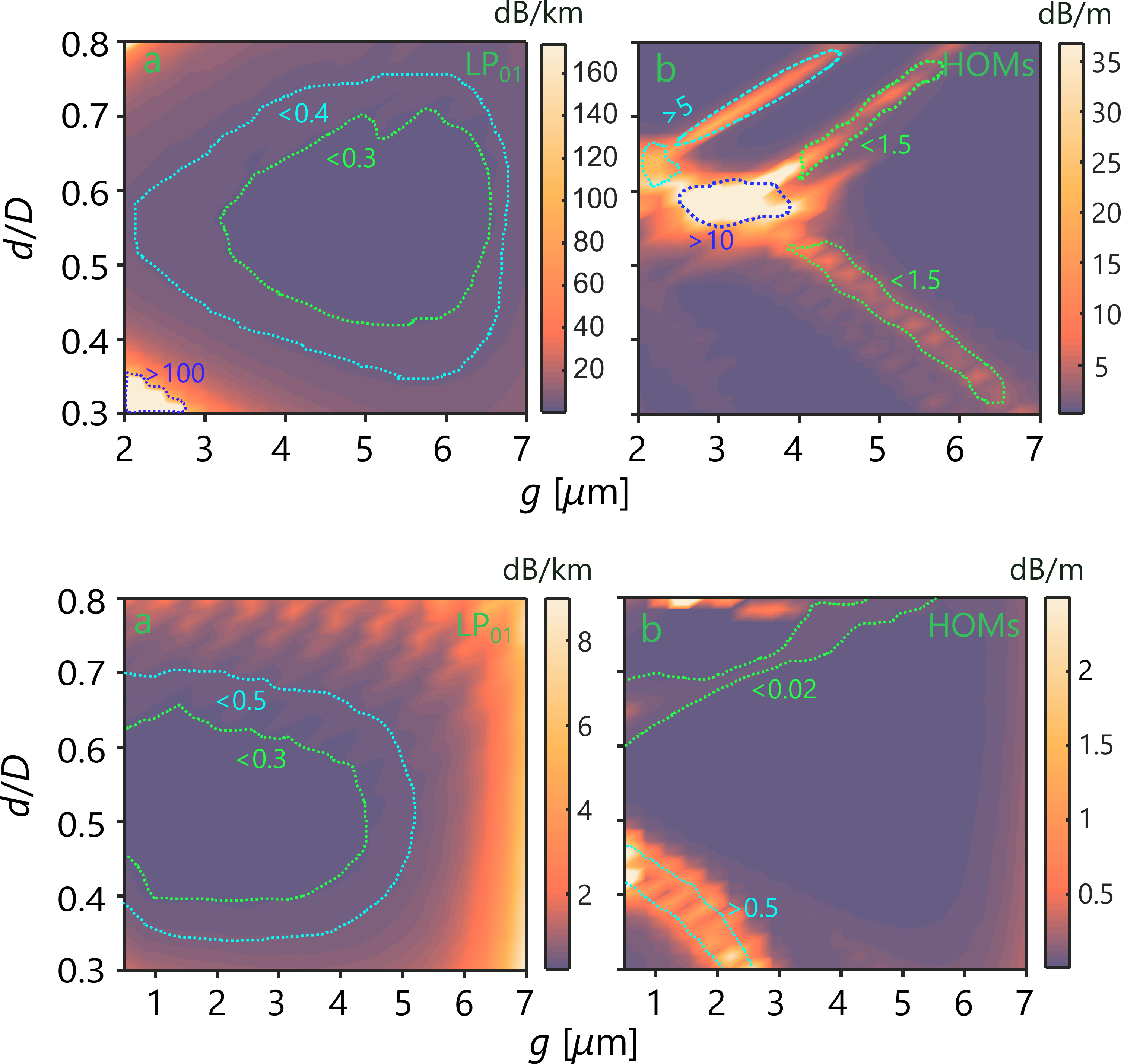}\\
  \caption{Calculated propagation loss of (a) LP$_{01}$-like FM and (b) HOMs as a function of $d/D$ with different values of gap separation, $g$ for 5-tube nested (top) and 6-tube nested (bottom) HC-ARF. HOMs loss is defined as the lowest loss among the four LP modes (LP$_{11a}$, LP$_{11b}$, LP$_{21a}$, and LP$_{21b}$). All simulations are performed at 1.55 $\mu$m. The fiber has a fixed core diameter, $D_\text{c}$ = 33 $\mu$m and average silica wall thickness, $t_1/t_2$ = 400/424 nm. To plot the 2D surface plots, gap separation, $g$, and normalized nested tube ratio, $d/D$ were scanned with 20 and 30 data points respectively and between the data points are interpolated. The colorbar for LP$_{01}$ and HOMs are in dB/km and dB/m respectively.}\label{fig:fig_9}
  \end{center}
\end{figure}

\subsection{Bend loss analysis}
Here, we investigate the effect of gap separation, $g$ and nested tube ratio, $d/D$ on bend loss. To calculate the bend loss, the bent structure is transformed into its equivalent straight structure with equivalent refractive index profile, $n_{\rm{eq}}$ defined by \cite{heiblum1975analysis} $n_{\text{eq}}=n(x,y)e^{(x,y)/R_\text{b}}$, where $R_\text{b}$ is the bending radius, $(x,y)$ is the transverse distance from the center of the fiber, $n(x,y)$ is the refractive index profile of the straight fiber. The elasto-optic effect was not included in our bend loss calculations because most of the light is guided inside the air-core whereas a little fraction of light leaks to the cladding \cite{poletti2014nested}. Figure~\ref{fig:fig_7} shows the bend loss of LP$_{01}$ (top) and HOMs (bottom) for a bend radius of 5 cm. The FM loss remains $<$1.5 dB/km in the range of 1.9 $\mu$m $<g< $ 5.9 $\mu$m and 0.42 $<d/D< $0.69 with a minimum loss of $<$0.7 dB/km. The loss trend is similar to what we observed in Fig.~\ref{fig:fig_5}. The FM loss increases for a small and large values of $d/D$. The HOMs can be highly phase-matched with CMs under bend condition by choosing suitable values of $g$ and $d/D$. The maximum HOMs loss of $\approx$ 5 dB/m was obtained for $g$ = 0.5 $\mu$m and $d/D$ = 0.6. Figure~\ref{fig:fig_8} shows the bend loss of an ideal and realistic 6-tube nested HC-ARF as a function of bend radius. In the small bend radii regime, the ideal structure has improved bend loss performance compared to the realistic one. However, both structures have comparable bend loss when the bend radii is large. Both fibers have bend loss of $<$1 dB/km for bend radius of $>$5 cm. It is interesting to note that despite of the large core dimensions, FM do not couple with CMs even with small bend radius. The results indicate that that the fiber can be bent with small bend radius.  

\subsection{Proposal for low-loss, wide bandwidth, and single-mode fiber design}
In this section, we discuss on the design of low-loss and wide bandwidth  HC-ARF design. In the previous designs mentioned in Sec. 3, micro-bend loss was the dominant loss factor in the short wavelength regime. In a recent experiments, it is mentioned that the impact of the micro-bend loss can be reduced and neglected by reducing the core size and use of improved coatings \cite{jasion2020hollow}. The micro-bending induced power loss is low for small core size in which the magnitude grows approximately $R_c^2$, where $R_c$ is the core radius \cite{fokoua2016microbending}. From the micro-bend loss equation in \cite{fokoua2016microbending}, it can be clearly seen that the micro-bend reduces with the decrease of core size. Therefore, in our proposed design, we used an HC-ARF with reduced core diameter of 33 $\mu$m, and $t_1/t_2$ = 400/424 nm (we scaled down the tube thickness accordingly compared to  the design in Fig.~\ref{fig:fig_1} and here the inner tube is 6$\%$ thicker than outer tube). A 5-tube nested HC-ARF was chosen because it offers wider transmission bandwidth and better single-mode performance compared to any number of tubes \cite{habib2019single}.

The 2D optimization of 5-tube nested HC-ARF is shown in Fig.~\ref{fig:fig_9} (top). The contribution of leakage loss, SSL, and material attenuation of silica was included to calculate the propagation loss whereas impact of micro-bend loss was neglected. The 5-tube nested HC-ARF offers low-loss over a wide range of $g$ and $d/D$ with a minimum loss of $\approx$0.25 dB/km. Unlike 6-tube nested HC-ARF (see Fig.~\ref{fig:fig_5}), the fiber has high loss for small values of $g$ and $d/D$. HOM loss was plotted by choosing the lowest loss among the four LP modes (LP$_{11a}$, LP$_{11b}$, LP$_{21a}$, and LP$_{21b}$). The 2D surface plot for HOMs shows that it follows similar trend as 6-tube nested HC-ARF (we called the shape of HOMs as `V-shape' pattern). We found that this kind of unique `V-shape' HOMs configuration can be generated irrespective of core diameters while maximum HOM loss occurs for $d/D\approx$0.6 \cite{habib2019single}. There is a wide range of $g$ and $d/D$ in which HOMs are highly suppressed. The maximum HOM loss of $>$35 dB/m was achieved whereas the FM loss remains 0.37 dB/km. The HOM loss is \emph{100K$\times$} higher than FM meaning that the fiber will not guide any HOM after propagating few meters. The HOM loss of the 5-tube nested HC-ARF is much higher than any number of tubes which confirms recent predictions \cite{habib2019single}. The SSL was calculated to be $\approx$0.09 dB/km at 1.55 $\mu$m. As mentioned, SSL was predicted according to \cite{roberts2005ultimate} in which SSL model was based on a different fiber type (\emph{i.e.,} HC-PBGF) and it was modeled and fitted without considering micro-bend and intermodal related losses \cite{jasion2020hollow}. Therefore, SSL might have negligible role for HC-ARF compared to HC-PBGF. Understanding the effect of SSL and micro-bend loss on the overall loss performance of this novel HC-ARF is still a debate and requires accurate modeling. However, this is beyond the scope of this work, we left this discussion for future investigations. It is worth to mention that, neglecting the effect of SSL will bring down the propagation loss of HC-ARF $<$0.17 dB/km at 1.55 $\mu$m close to the loss level of SSMF \cite{tamura2018first}. 

\begin{figure}
  \begin{center}
  \includegraphics[width=3.7in]{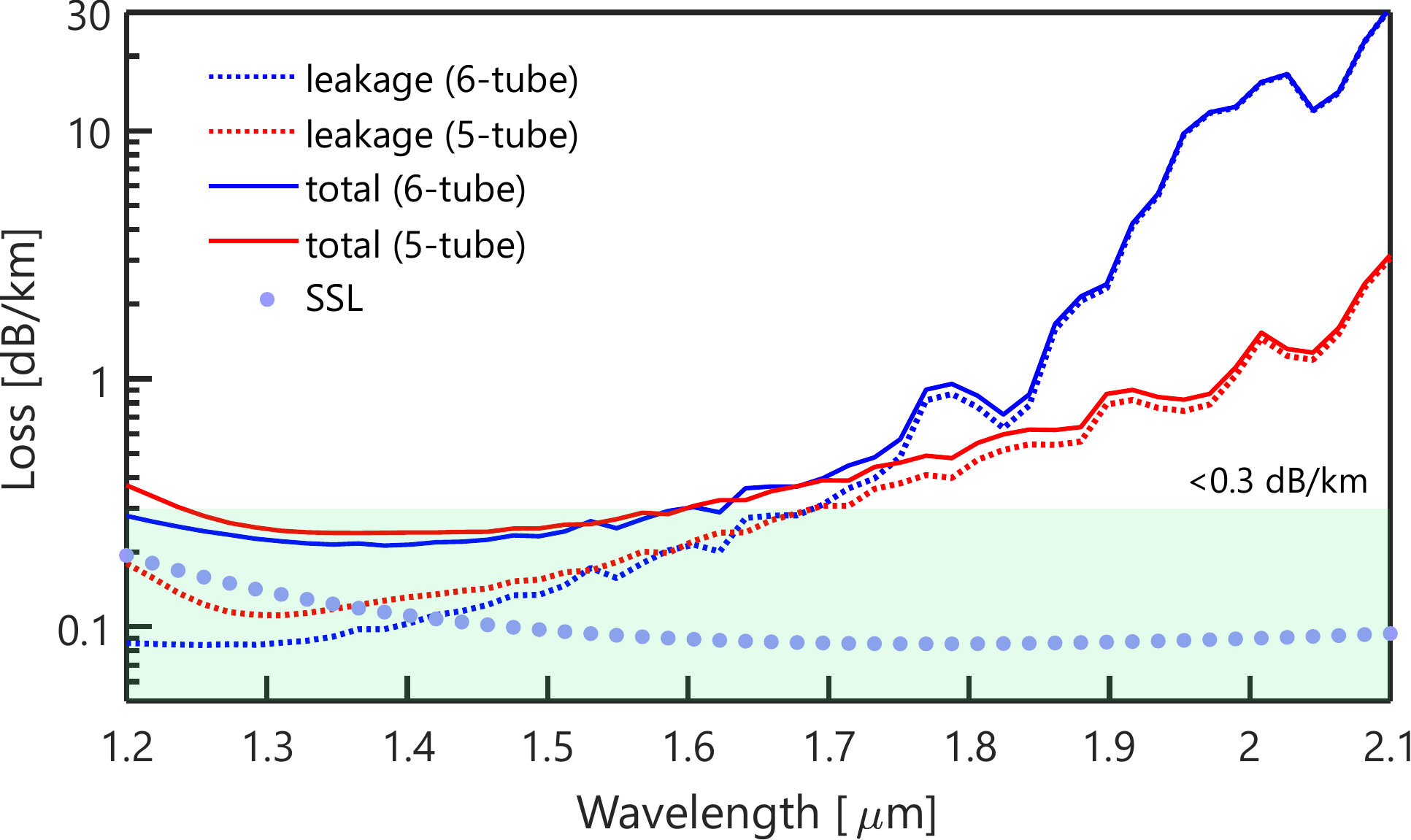}\\
   \caption{Calculated optimized propagation loss spectra of 5-tube and 6-tube nested HC-ARF with gap separations, $g$ = 5.1 $\mu$m and $g$ = 3 $\mu$m, normalized tube ratio, $d/D$ = 0.52 and $d/D$ = 0.5 respectively. The fibers have core diameter, $D_\text{c}$ = 33 $\mu$m and average silica wall thickness, $t_1/t_2$ = 400/424 nm. The micro-bend loss was not included in the overall loss calculation. solid lines: propagation loss, broken lines: leakage loss, and closed circles: SSL.}\label{fig:fig_10}
  \end{center}
\end{figure}

As a comparison, 2D surface plot of 6-tube nested HC-AR fiber is illustrated in Fig.~\ref{fig:fig_9} (bottom). The propagation loss of FM and HOMs follows similar pattern as shown in Fig.~\ref{fig:fig_5}. The low loss region is smaller compared to 5-tube nested HC-ARF. The HOMs loss pattern seems to be `V-shaped'. However, the HOMs loss is far lower than the 5-tube nested HC-ARF. The maximum loss of HOMs is found just above 2 dB/m while the FM loss is slightly above 0.5 dB/km. A propagation loss comparison between a 5-tube and 6-tube nested HC-ARF is shown in Fig.~\ref{fig:fig_10}. The 6-tube HC-ARF has slightly better performance than 5-tube HA-ARF in 1.2--1.7 $\mu$m wavelength range. However, in the longer wavelength regime, 5-tube nested HC-ARF has much lower loss compared to 6-tube nested HC-ARF. This is due to the fact that 5-tube HAC-ARF has larger distance between the core surround and the outer tube compared to 6-tube HC-ARF.

\begin{figure}
  \begin{center}
  \includegraphics[width=4.5in]{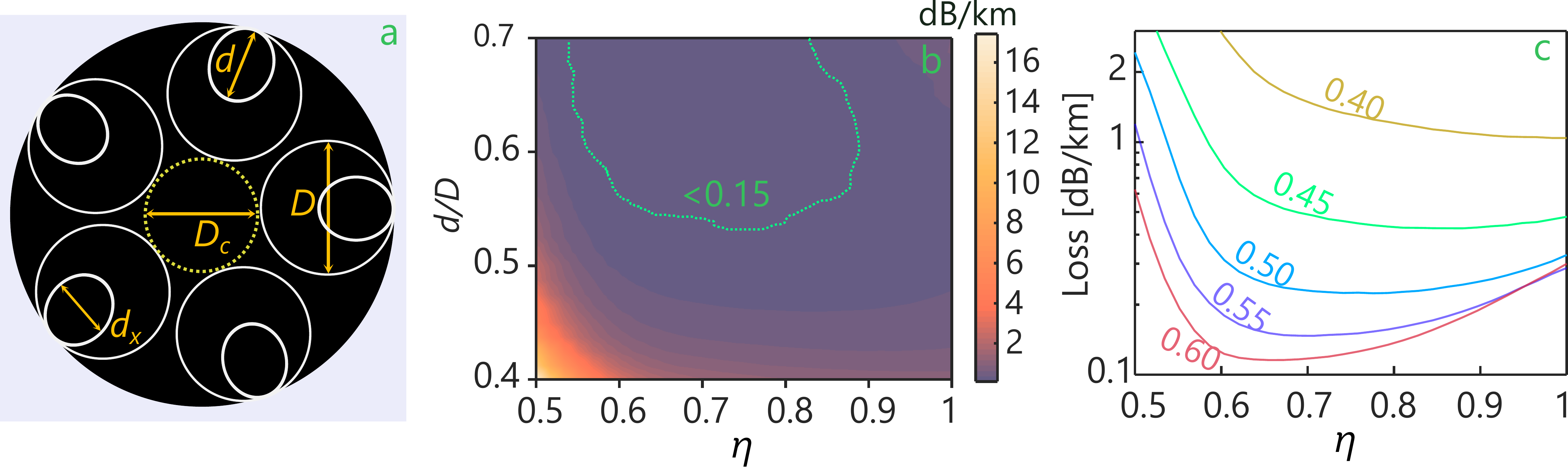}\\
  \caption{(a) Geometry of a 5-tube anisotropic nested HC-ARF in which major and minor axis is denoted by $d$ and $d_x$ respectively. Calculated propagation loss of LP$_{01}$-like FM as a function of ellipticity, $\eta$ (b) with different $d/D$ and (c) for 5 values of $d/D$ for 5-tube anisotropic nested HC-ARF. All simulations are performed at 1.55 $\mu$m. The fiber has a fixed core diameter, $D_\text{c}$ = 33 $\mu$m, gap separation, $g$ = 4 $\mu$m, and average silica wall thickness, $t_1/t_2$ = 400/424 nm. To plot the 2D surface plot normalized tube ratio, $d/D$, and ellipticity, $\eta$ were scanned with 25 and 30 data points respectively and between the data points are interpolated.}\label{fig:fig_11}
  \end{center}
\end{figure}

Next, we investigate the performance of the 5-tube nested HC-ARF with anisotropic (elliptical) nested anti-resonant tubes while outer tubes remain circular. Figure~\ref{fig:fig_11}(a) shows the geometry of a 5-tube anisotropic nested HC-ARF. Introducing such anisotropic nested tubes allow increased negative curvature to the core which allows strong confinement of light in the core. The ellipticity is defined as $\eta = d_x/d$, where $d_x$ is the diameter in the azimuthal direction and $d$ is the diameter in the radial direction. The 2D optimization for FM loss of 5-tube anisotropic nested HC-ARF at 1.55 $\mu$m is shown in Fig.~\ref{fig:fig_11}(b). It can clearly be seen from Fig.~\ref{fig:fig_11}(b) that there is a wide range of $d/D$ and $\eta$ in which the propagation loss $<$0.15 dB/km which is indicated by broken green lines with a minimum loss $<${0.11} dB/km. The propagating loss increases with decreasing values of $d/D$ and $\eta$. To keep the propagation loss $<$0.15 dB/km, nested tube ratio $d/D$ needs to maintain $>$0.55. Fig.~\ref{fig:fig_11}(c) shows that propagation loss strongly depends on $d/D$.

\begin{figure}
  \begin{center}
  \includegraphics[width=3.68in]{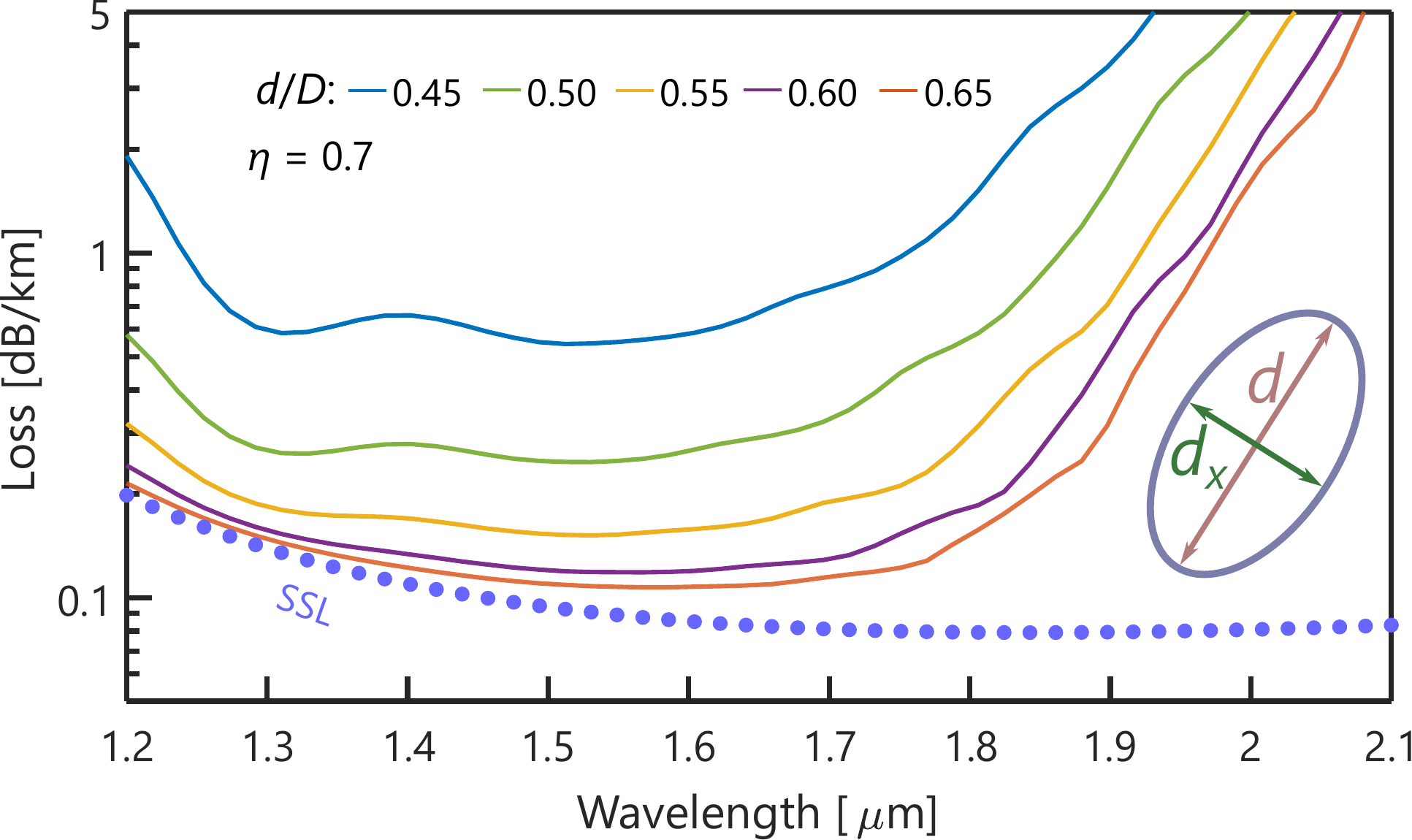}\\
  \caption{Calculated propagation loss spectra of 5-tube anisotroptic nested HC-ARF for a fixed ellipticity of $\eta$ = 0.7 and different values of $d/D$. The fiber has a fixed core diameter, $D_\text{c}$ = 33 $\mu$m, gap separation, $g$ = 4 $\mu$m, and average silica wall thickness, $t_1/t_2$ = 400/424 nm.}\label{fig:fig_12}
  \end{center}
\end{figure}

The propagation loss spectra for a fixed $\eta$ of 0.7 and different $d/D$ is illustrated in Fig.~\ref{fig:fig_12}. Since the anisotropic HC-ARF performs well over a broad range of ellipticity, we choose a moderate value of ellipticity $\eta$ = 0.7. The predicted SSL was around 0.08--0.20 dB/km in the wavelength range of 1200--2100 nm with $\approx$0.09 dB/km at 1.55 $\mu$m. The minimum calculated leakage loss was found around $<$0.02 dB/km at 1.55 $\mu$m for $d/D$ = 0.65. The leakage loss has negligible contribution while SSL dominates over leakage loss. It is seen from Fig.~\ref{fig:fig_12} that propagation loss strongly depends on $d/D$, and decreases with increasing $d/D$. The optimized design offers propagation loss as low as 0.11 dB/km at 1.55 $\mu$m, which we believe this design can potentially beat the loss of SSMF \cite{tamura2018first}. The propagation loss remain $<$0.2 dB/km in the wavelength range of 1200--1850 nm.

\section{Conclusion}
In summary, we have investigated the effect of the structure of nested tubes on the overall loss performance of HC-ARF, in our analysis by allowing the nested tubes to penetrate more into the outer jacket.
Importantly, we found that in order to achieve low loss of regular HC-ARF, the size of the nested tubes is more crucial compared to their precise shape. In fact, when the geometry of the nested tubes become non-ideal (such scenario is likely to come up in the manufacturing process), the fiber performance remains almost unchanged compared to that of an ideal structure by slightly inflating the non-ideal tubes. 
Next, we have identified strongly HOM suppression region which follows a `V-shape' pattern found by optimizing the gap separation, $g$ and nested tube ratio, $d/D$. Furthermore, A 5-tube nested HC-ARF design was proposed which exhibits propagation loss of $<$0.25 dB/km at 1.55 $\mu$m, wide transmission window, and truly single-mode operation. Finally, we study HC-ARF with anisotropic (elliptical) nested tubes which significantly reduces propagation loss close to 0.11 dB/km at 1.55 $\mu$m. We believe that exploring new capillary structures is valuable as this can open new design avenues - even if elliptical capillaries are not realistic with current fabrication techniques. We expect that our results will give a better insight for designing single-mode, wide-band, and ultra low-loss HC-ARFs for multitude of applications including high speed optical communication, high energy light transport, and ultrafast non-linear optics. 

\section*{Funding}
Army Research Office (ARO) (W911NF-17-1-0501 and W911NF-12-1-0450); Air Force Office
of Scientific Research (AFOSR) FA9550-15-10041); Danish Research Council (4184-
00359B and 8022-00091B).

\section*{Acknowledgments}
The authors would like to thank Francesco Poletti and Walter Belardi for useful discussions.

\section*{Disclosures}
The authors declare no conflicts of interest.

\bibliography{sample}






\end{document}